\documentclass[mathleft
]{an}
\usepackage{graphicx}
\usepackage{times}
\begin{document}


\title{Cloud Modeling of a Network Region in H$\alpha$ }

\author{Z. F. Bostanc\i \inst{1}\fnmsep\thanks{Corresponding author:
  \email{fbostanci@gmail.com}\newline}
}
\titlerunning{Cloud Modeling of a Network Region in H$\alpha$}
\authorrunning{Z.F. Bostanc\i}
\institute{
Istanbul University, Faculty of Sciences, Department of Astronomy and Space
Sciences, 34119 University, Istanbul, Turkey
}



\abstract{%
  In this paper, we analyze the physical properties of dark mottles in
  the chromospheric network using two dimensional spectroscopic
  observations in H$\alpha$ obtained with the G\"ottingen Fabry-Perot
  Spectrometer in the Vacuum Tower Telescope at the Observatory del
  Teide, Tenerife.  Cloud modeling was applied to measure the mottles'
  optical thickness, source function, Doppler width, and line of sight
  velocity. Using these measurements, the number density of hydrogen
  atoms in levels 1 and 2, total particle density, electron density,
  temperature, gas pressure, and mass density parameters were
  determined with the method of Tsiropoula \& Schmieder (1997). We
  also analyzed the temporal behaviour of a mottle using cloud
  parameters. Our result shows that it is dominated by 3 minute signals
  in source function, and 5 minutes or more in velocity. }

\maketitle

\section{Introduction}
The main fine structures of the quiet chromosphere emanating from the
network are called mottles when seen on the disk and spicules when
seen at the limb. Mottles appear as dark or bright structures, which
together form two kinds of groups, namely rosette and chain.  The
mottles in a rosette spread radially outwards from a bright center,
while in a chain they all point in the same direction. The dark
mottles are thin and elongated structures, whereas the bright mottles
are small and roundish, lying at a lower height than the dark
ones. These structures presumably extend along the magnetic fields and
cover between $7''$ and $10''$ with widths smaller than $1''$ and have
lifetimes of the order of 10 min or more.  The spatial relationship
between the bright and dark mottles is given in detail by Zachariadis
et al. (2001) (see also references therein).

There are a number of studies on the physical properties of
chromospheric fine structures using spectrometric observations as well
as various theoretical models and simulations (Sterling 2000;
Tziotziou et al. 2003; Al et al. 2004; De Pontieu et al. 2004;
Rouppe van der Voort et al. 2007).  Assuming an optically thin
absorbing cloud in front of a radiation source, the standard cloud
model (Beckers 1964) has been extensively used to extract intrinsic
line formation parameters such as source function, optical thickness,
Doppler width and line of sight velocity (see review by
Tziotziou 2007).

In this paper, we analyzed chromospheric features observed in
H$\alpha$ line with very good spectral profile sampling to derive
the features' physical properties in terms of standard cloud modeling.  Using
the method of Tsiropoula \& Schmieder (1997), additional physical
parameters were also derived from the inferred cloud model parameters.
We also addressed the dynamics of a dark mottle by the analysis of
cloud velocity and source function variations.  Our goal is to define
the physical conditions inside dark mottles and to investigate the
oscillatory behaviour by taking advantage of high resolution imaging
spectroscopic capabilities of the G\"ottingen Fabry-Perot spectrometer
and comparing our results with previous studies.

\section{Observations and Data Analysis}

The data were obtained with the Vacuum Tower Telescope (VTT)'s
G\"ottingen Fabry-Perot spectrometer, which is based on two
Fabry-Perot Interferometers (Koschinsky et al. 2001).  In 2002, a
short time series of 60 wavelength scans of the H$\alpha$ line were
taken from a network region near the disk center of the sun.  A
spacing of 125\,m\AA\, between adjacent wavelength positions was
chosen for the spectral scans. With the narrow-band channel, 8 images
at each of 18 wavelength positions were obtained, which provides a
wavelength coverage of approximately 2.125\,\AA. The exposure time was
30\,ms and the time interval between the start of two subsequent
spectral scans was 49\,s. The observed field of view of the raw data
was $38''.4\times28''.6$, with a spatial scale of $0''.1$ per pixel.
During the observations obtained under good seeing conditions,
broad-band images were taken simultaneously with narrow-band images.
Our best scan reached a Fried parameter of r$_{0}$=21\,cm.  Dark scans,
flat fields and scans with the continuum source were also taken for
the data reduction.

\begin{figure*}
\centering
\includegraphics[bb=100 63 475 290, clip,width=0.273\linewidth]{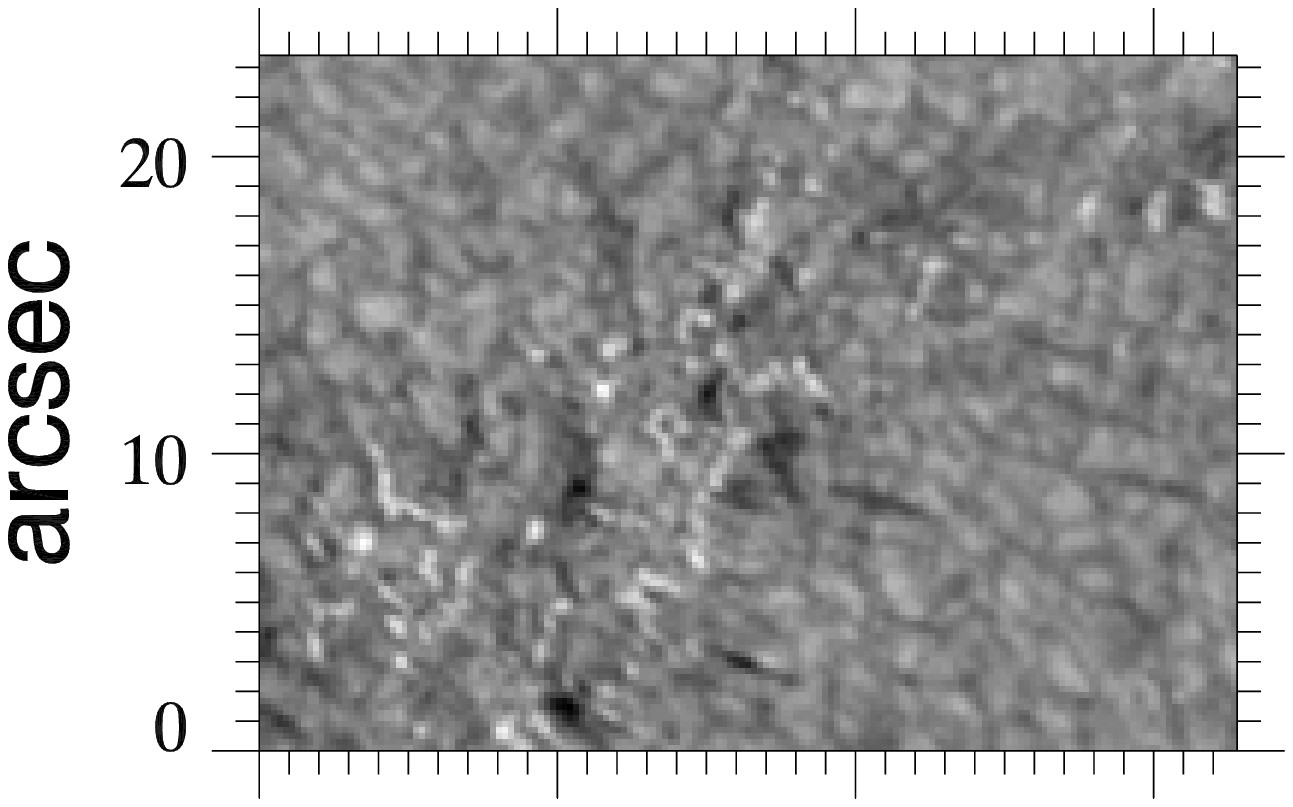}
\includegraphics[bb=100 61 475 318, clip,width=0.237\linewidth]{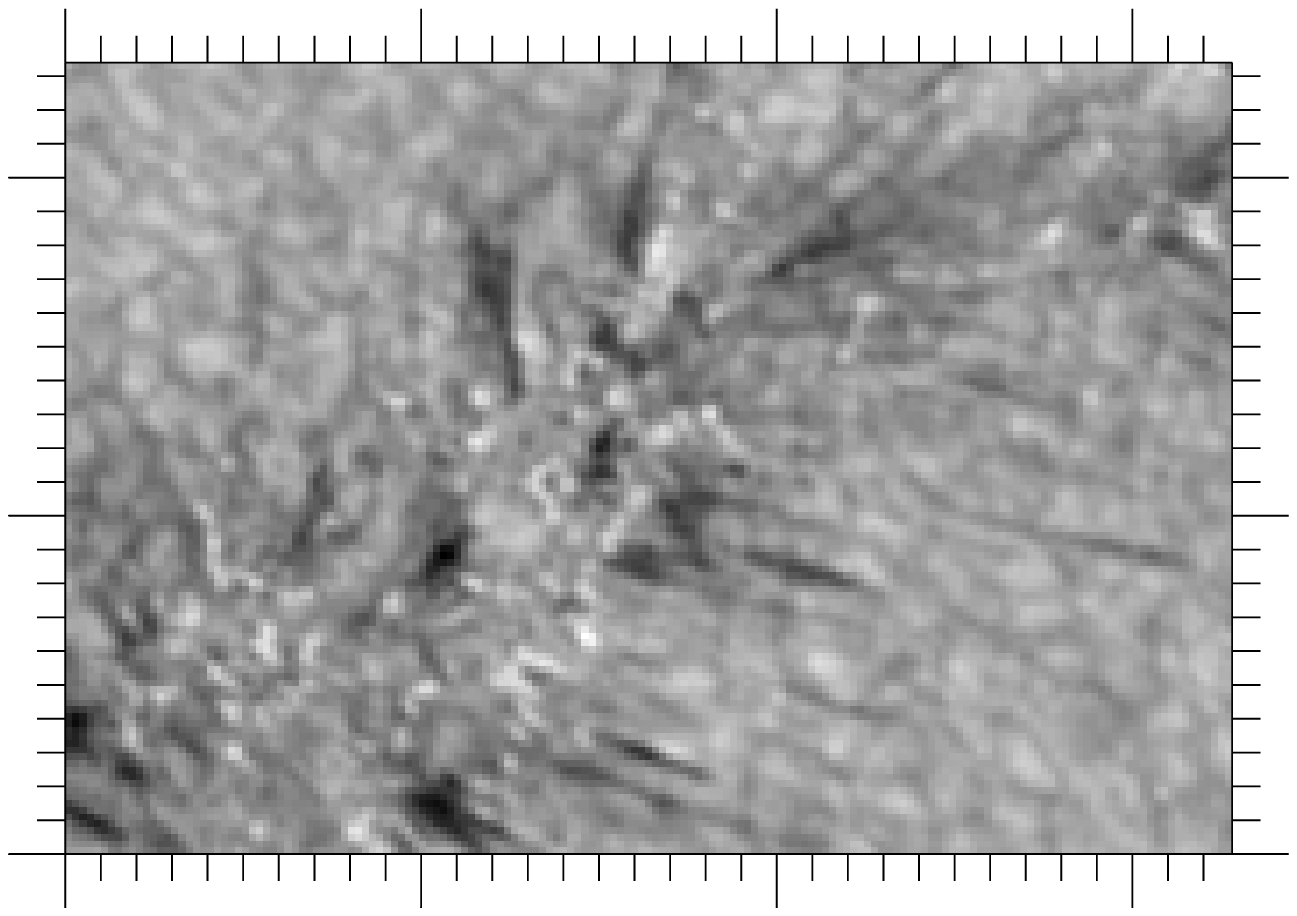}
\includegraphics[bb=100 61 475 318, clip,width=0.237\linewidth]{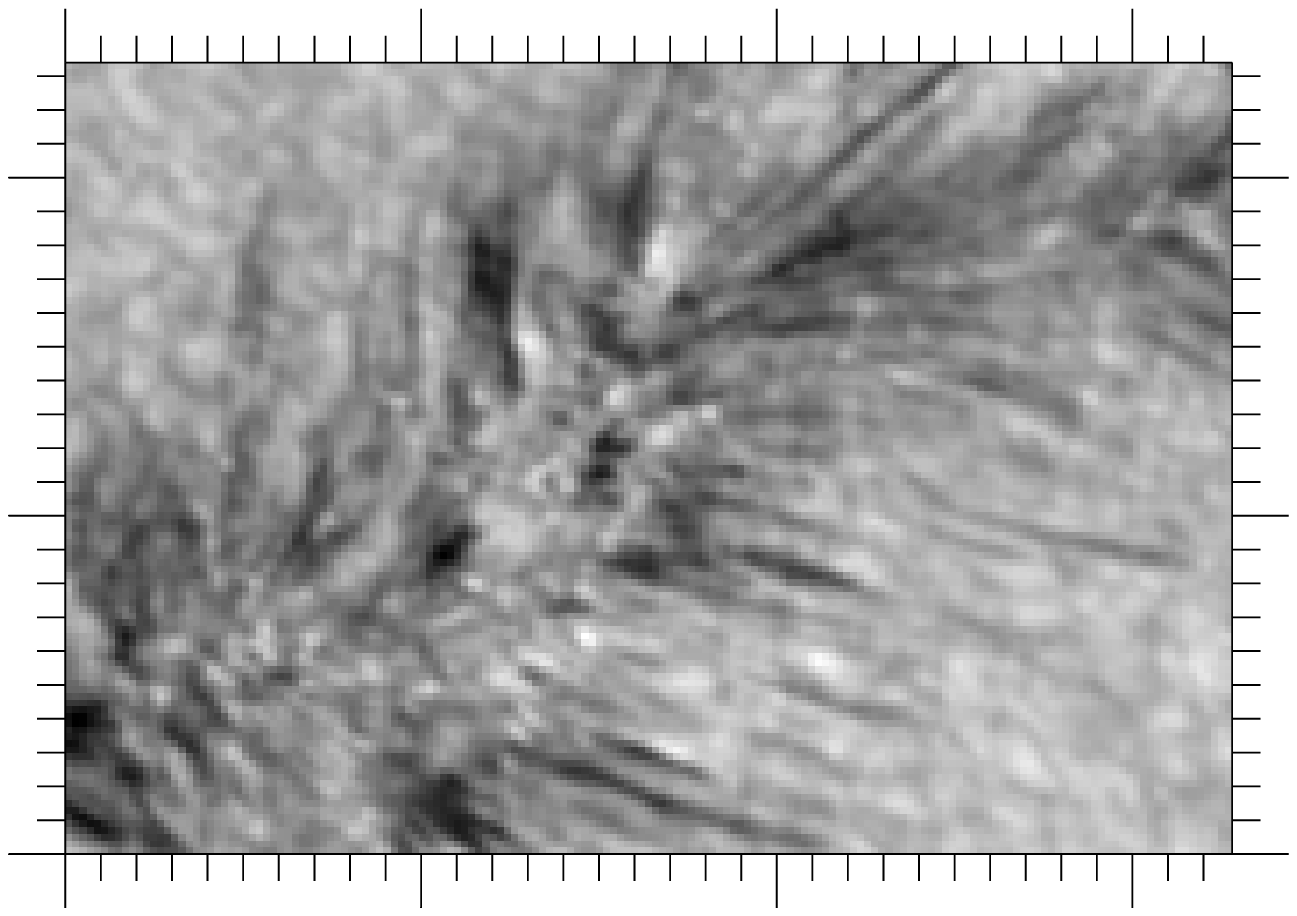}
\includegraphics[bb=100 61 475 318, clip,width=0.237\linewidth]{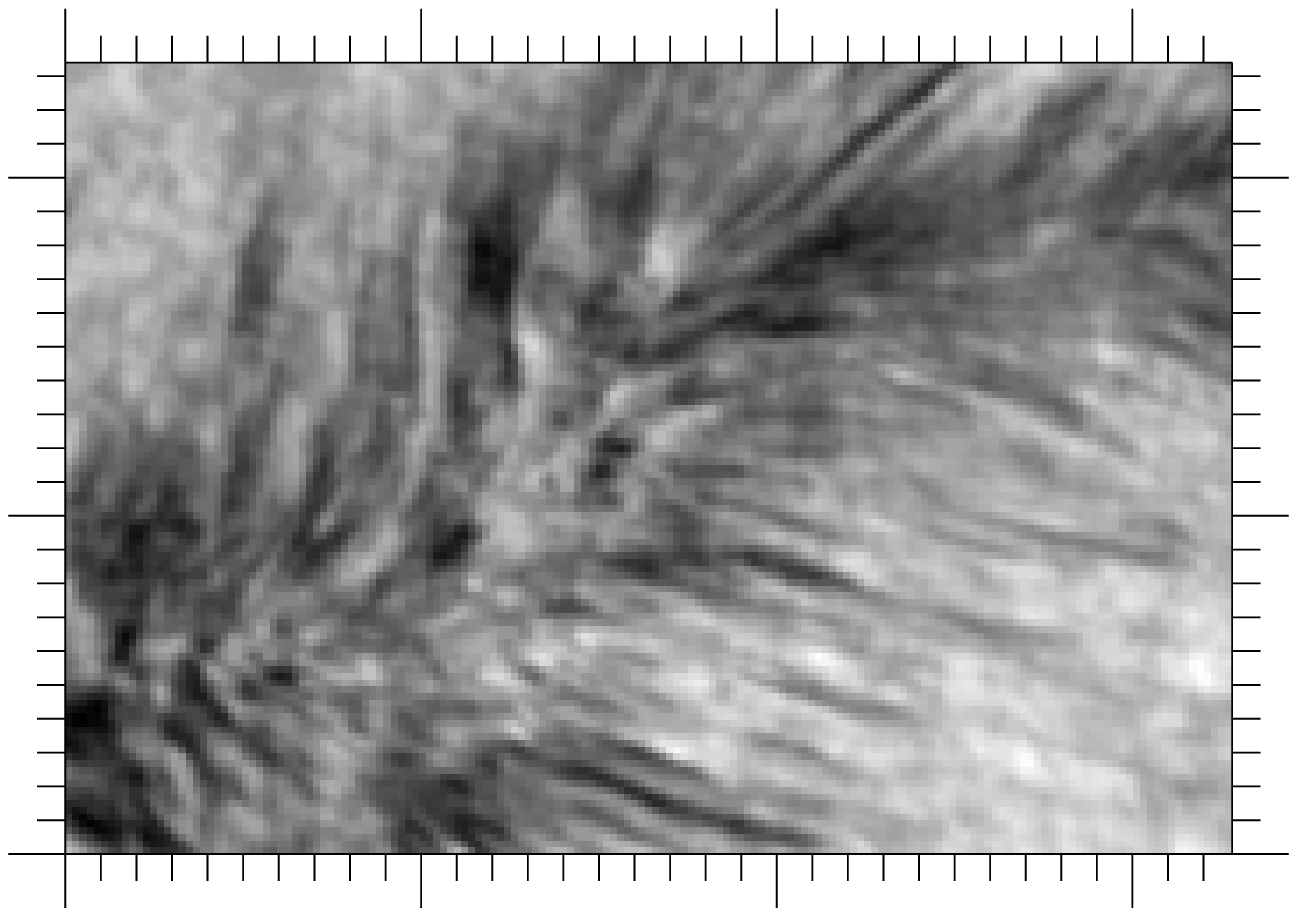}
\includegraphics[bb=100 63 475 290, clip,width=0.273\linewidth]{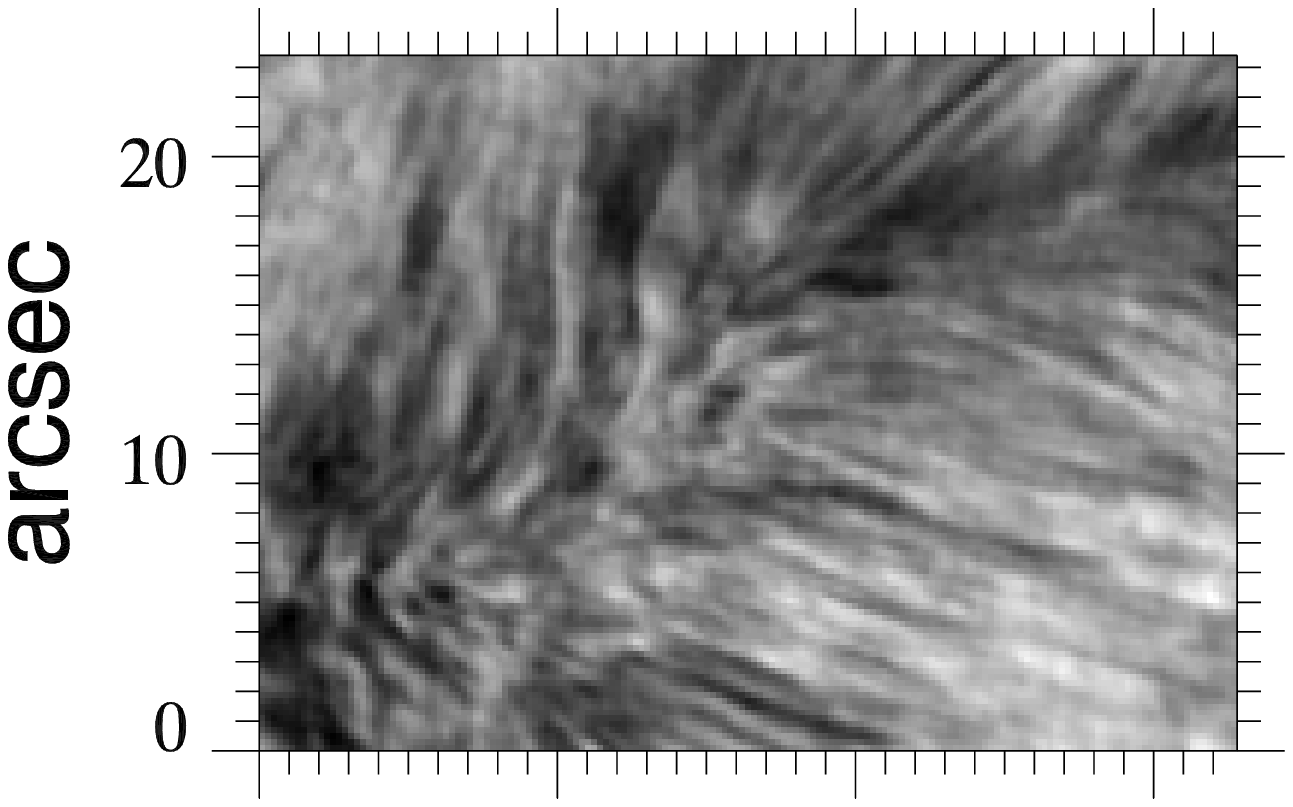}
\includegraphics[bb=100 61 475 318, clip,width=0.237\linewidth]{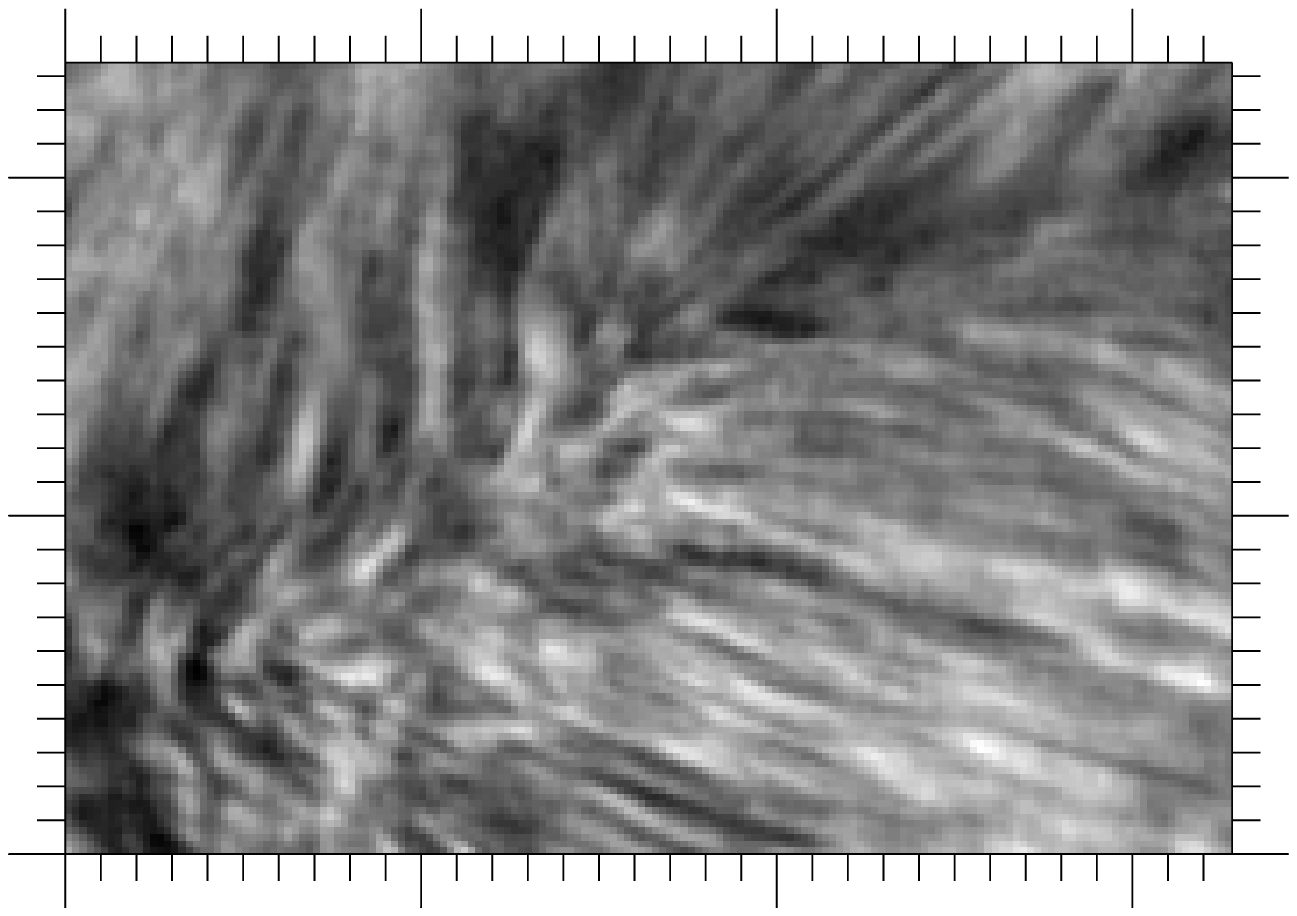}
\includegraphics[bb=100 61 475 318, clip,width=0.237\linewidth]{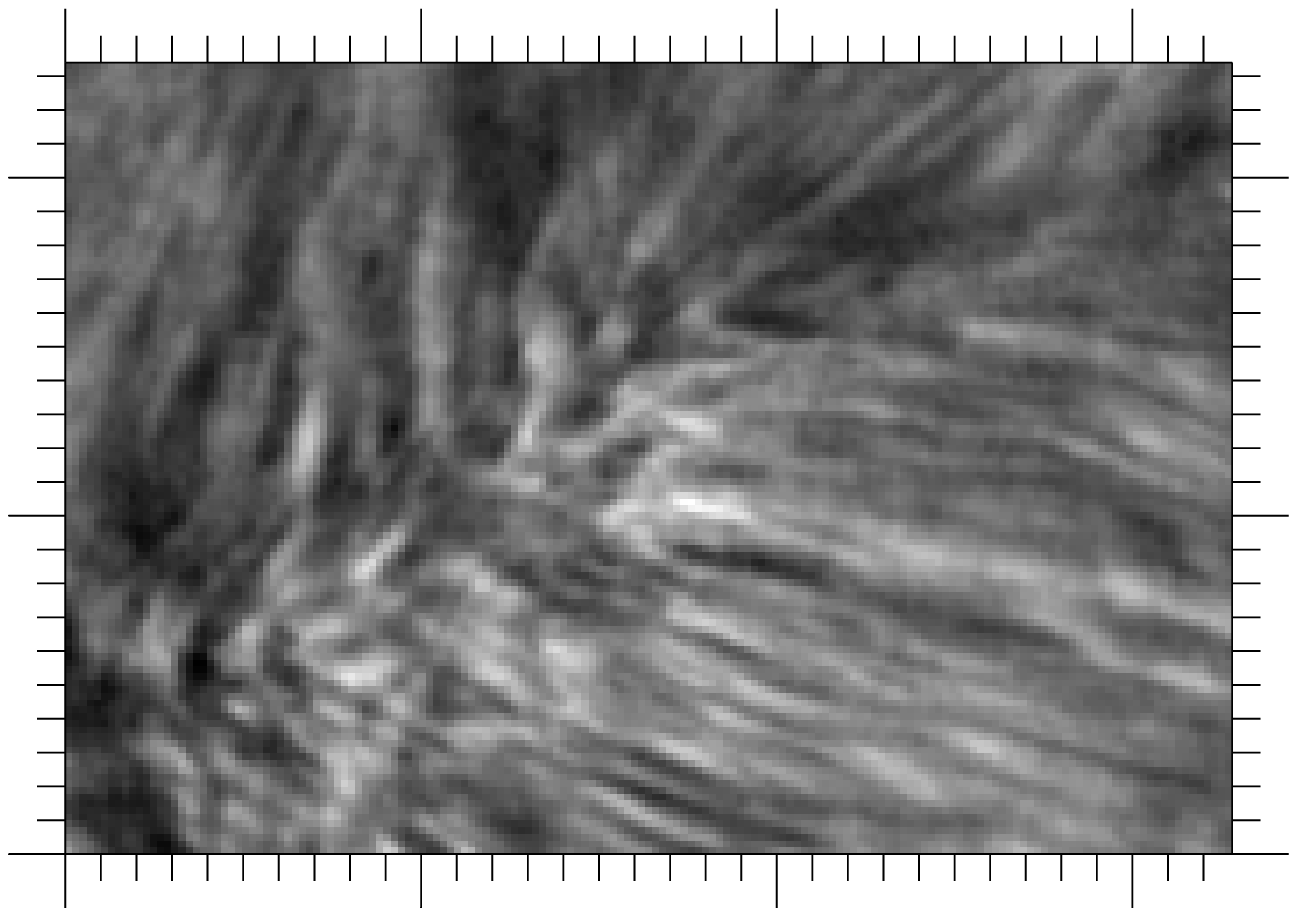}
\includegraphics[bb=100 61 475 318, clip,width=0.237\linewidth]{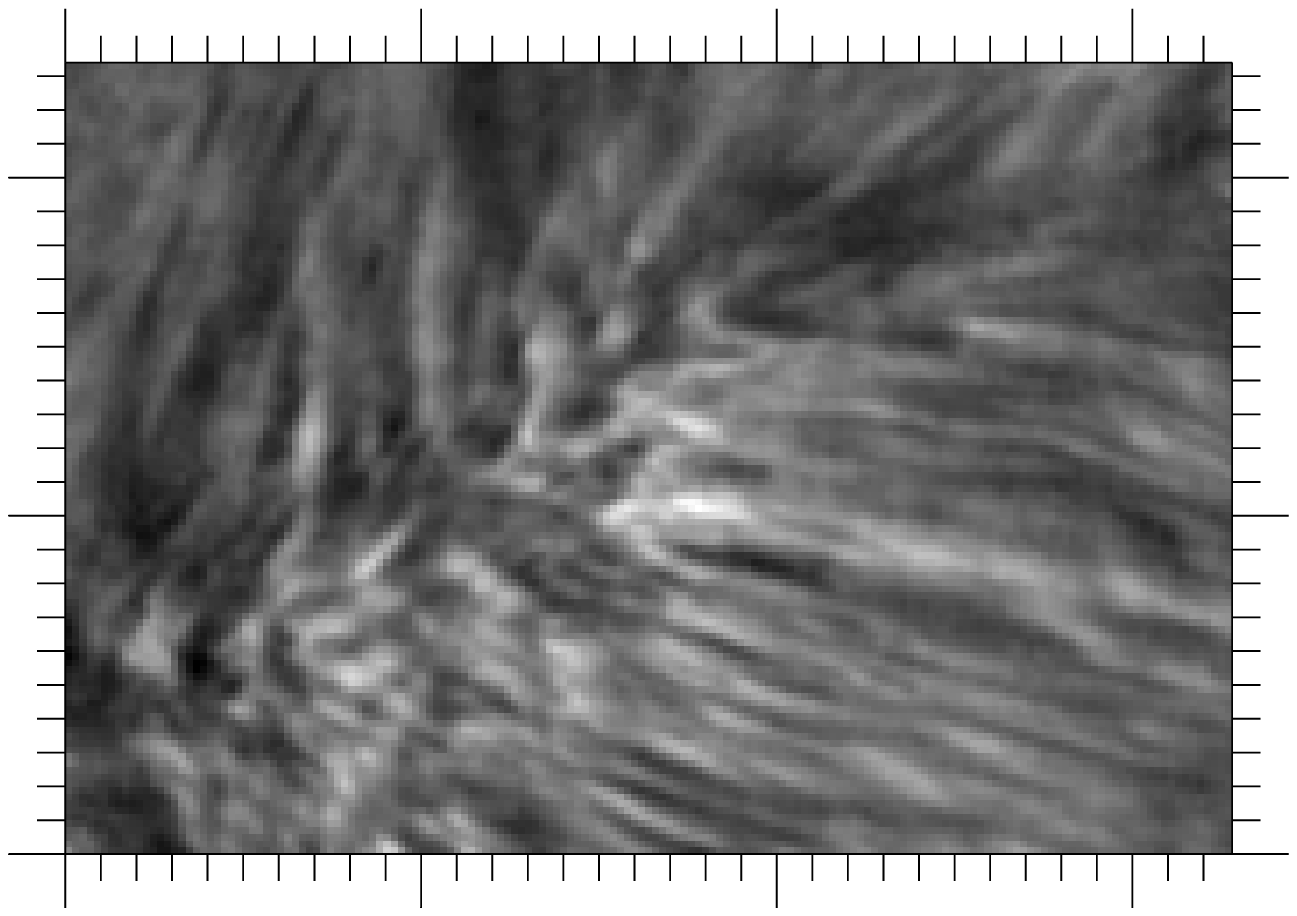}
\includegraphics[bb=100 63 475 290, clip,width=0.273\linewidth]{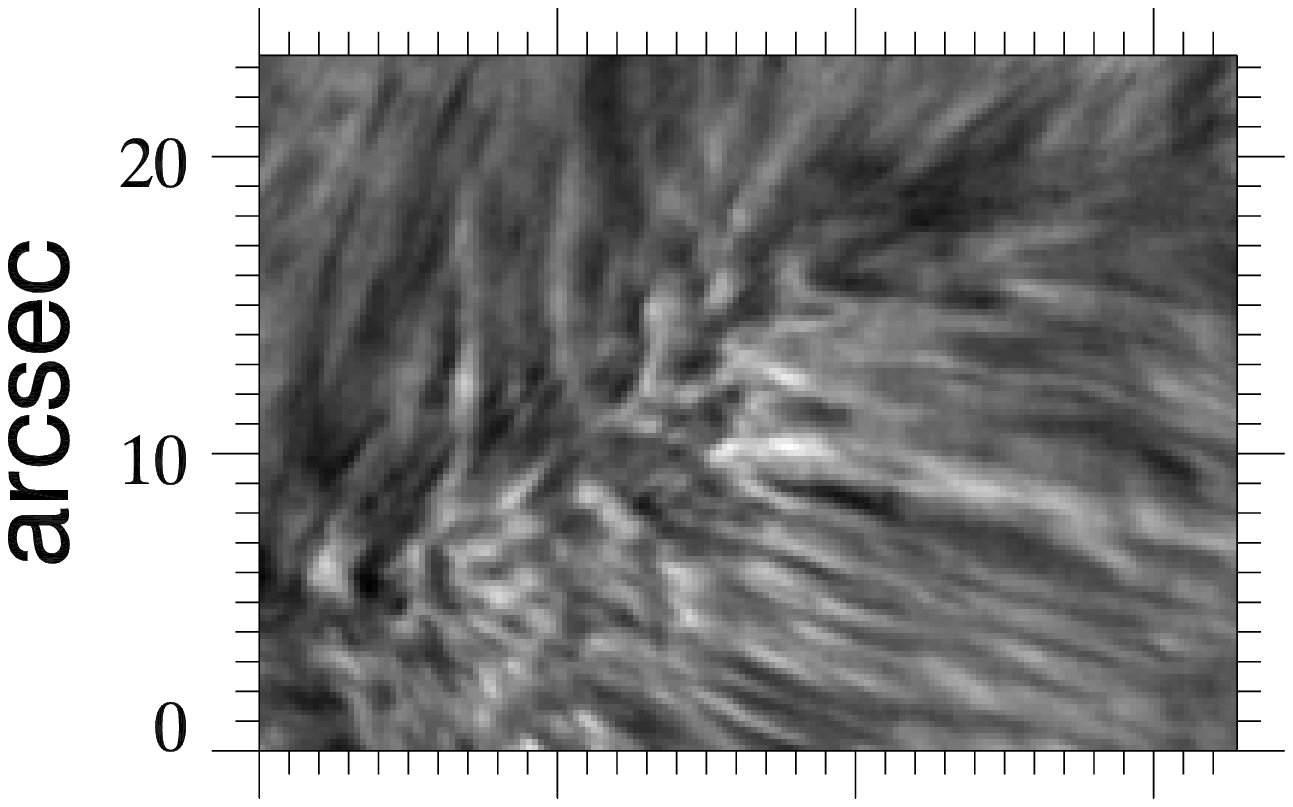}
\includegraphics[bb=100 61 475 318, clip,width=0.237\linewidth]{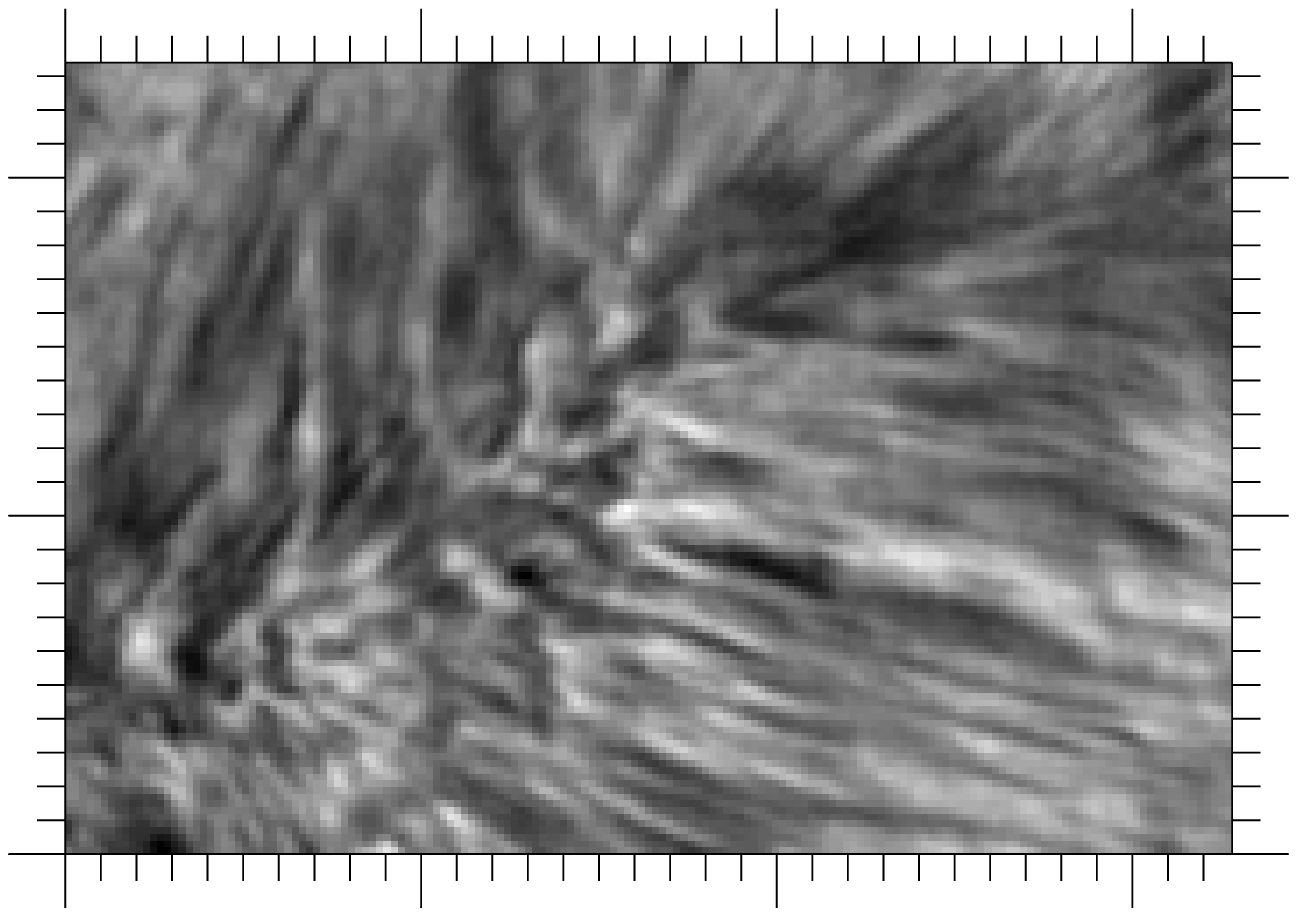}
\includegraphics[bb=100 61 475 318, clip,width=0.237\linewidth]{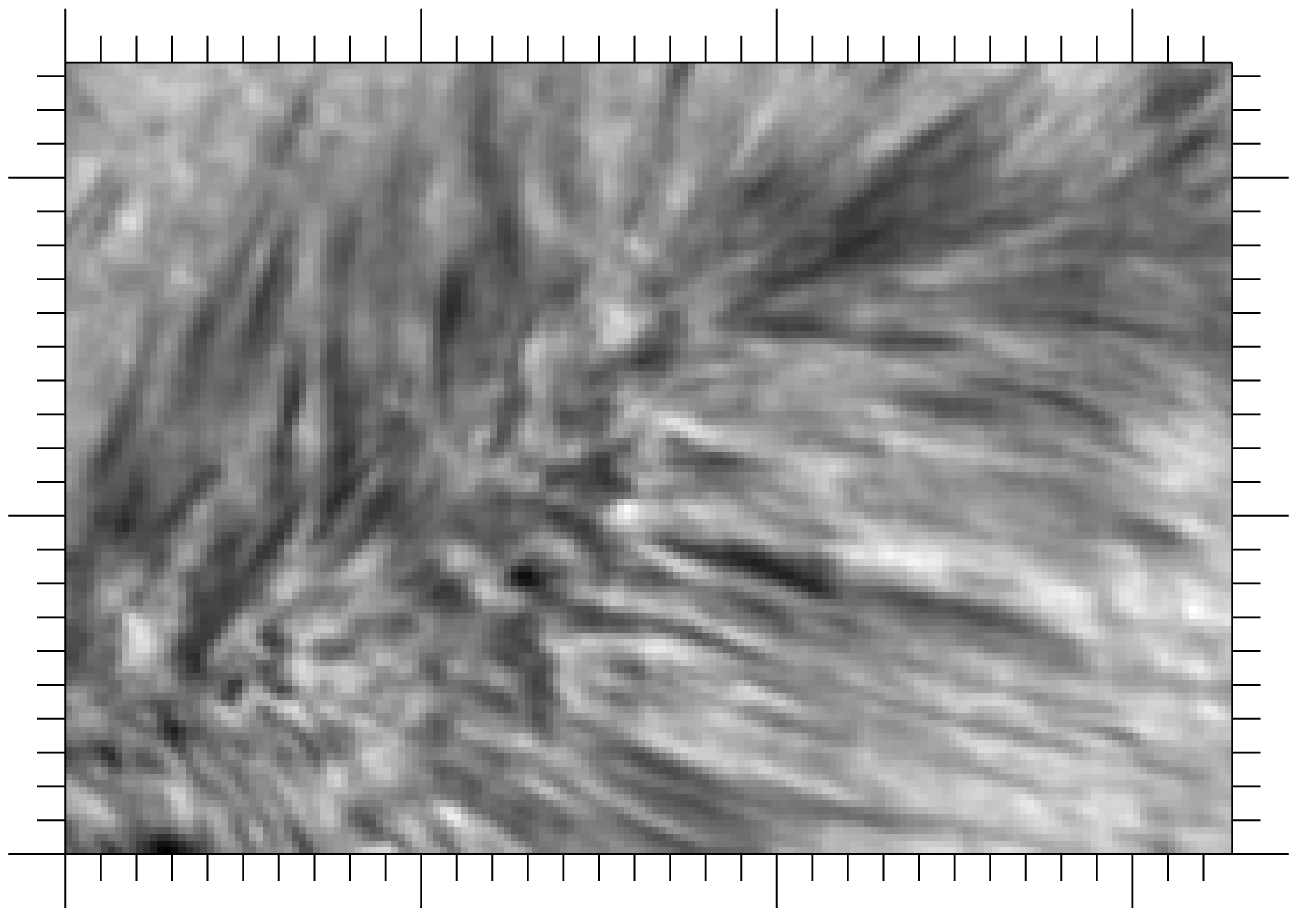}
\includegraphics[bb=100 61 475 318, clip,width=0.237\linewidth]{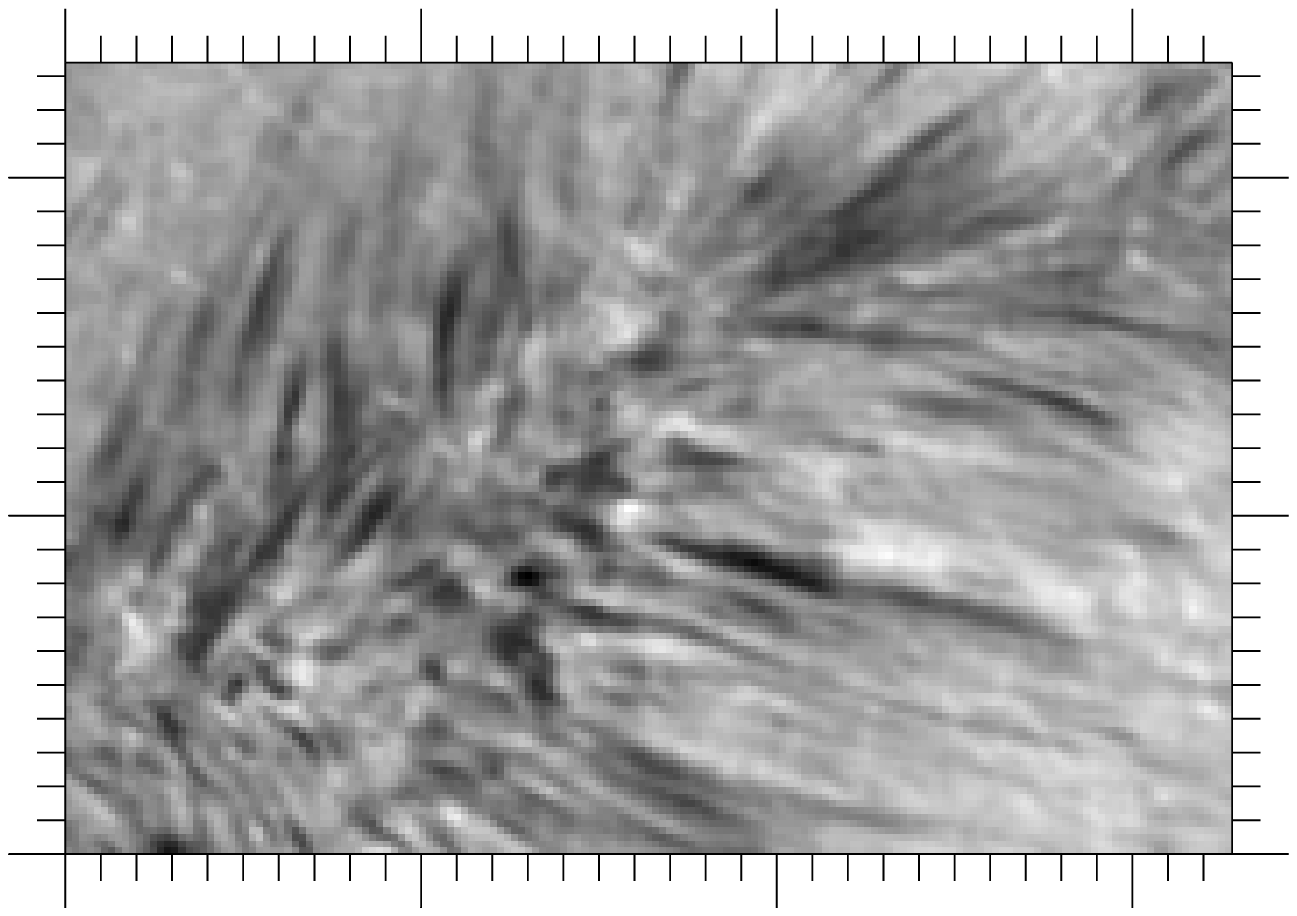}
\includegraphics[bb=100 61 475 360, clip,width=0.274\linewidth]{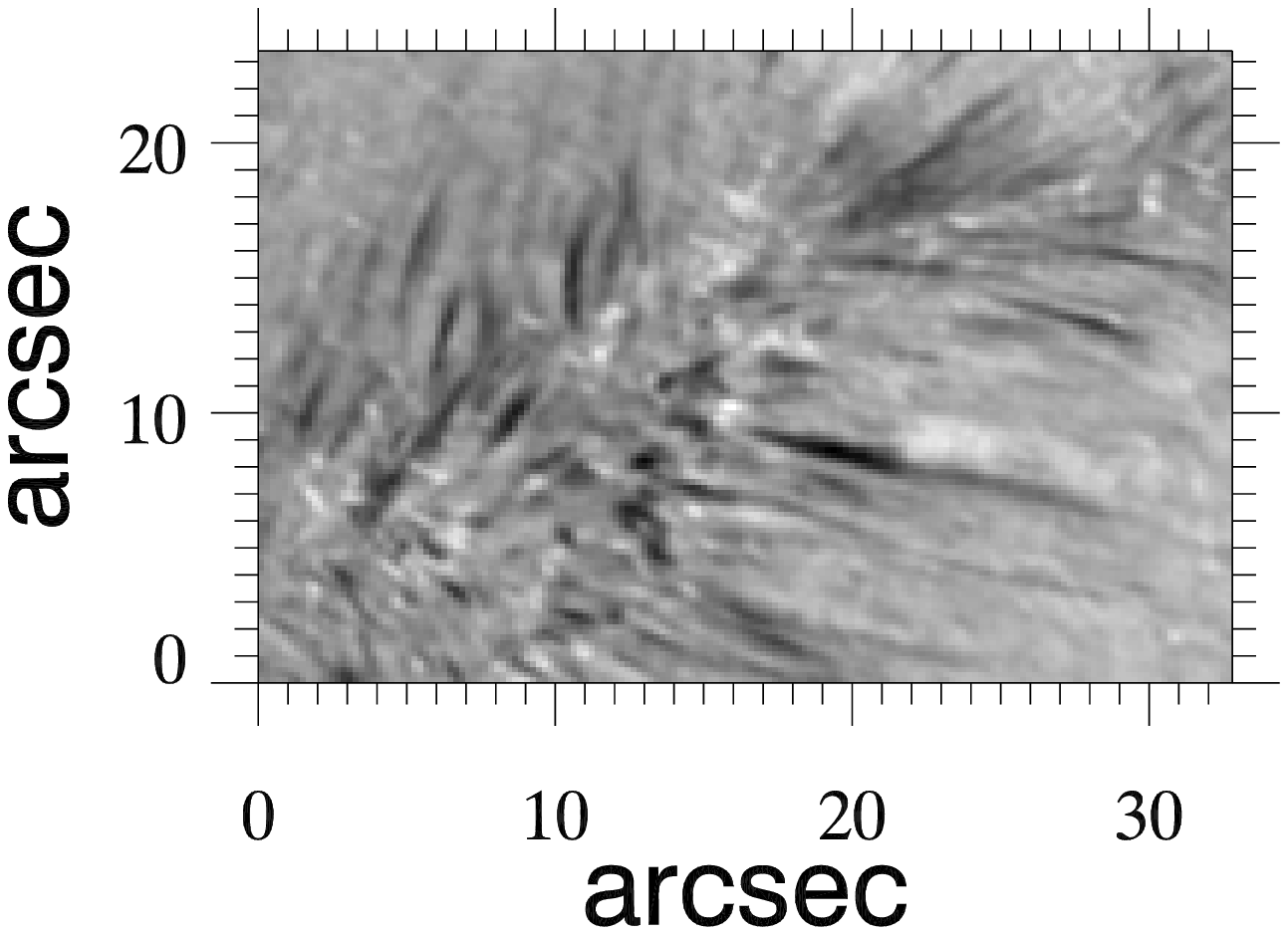}
\includegraphics[bb=100 61 475 400, clip,width=0.237\linewidth]{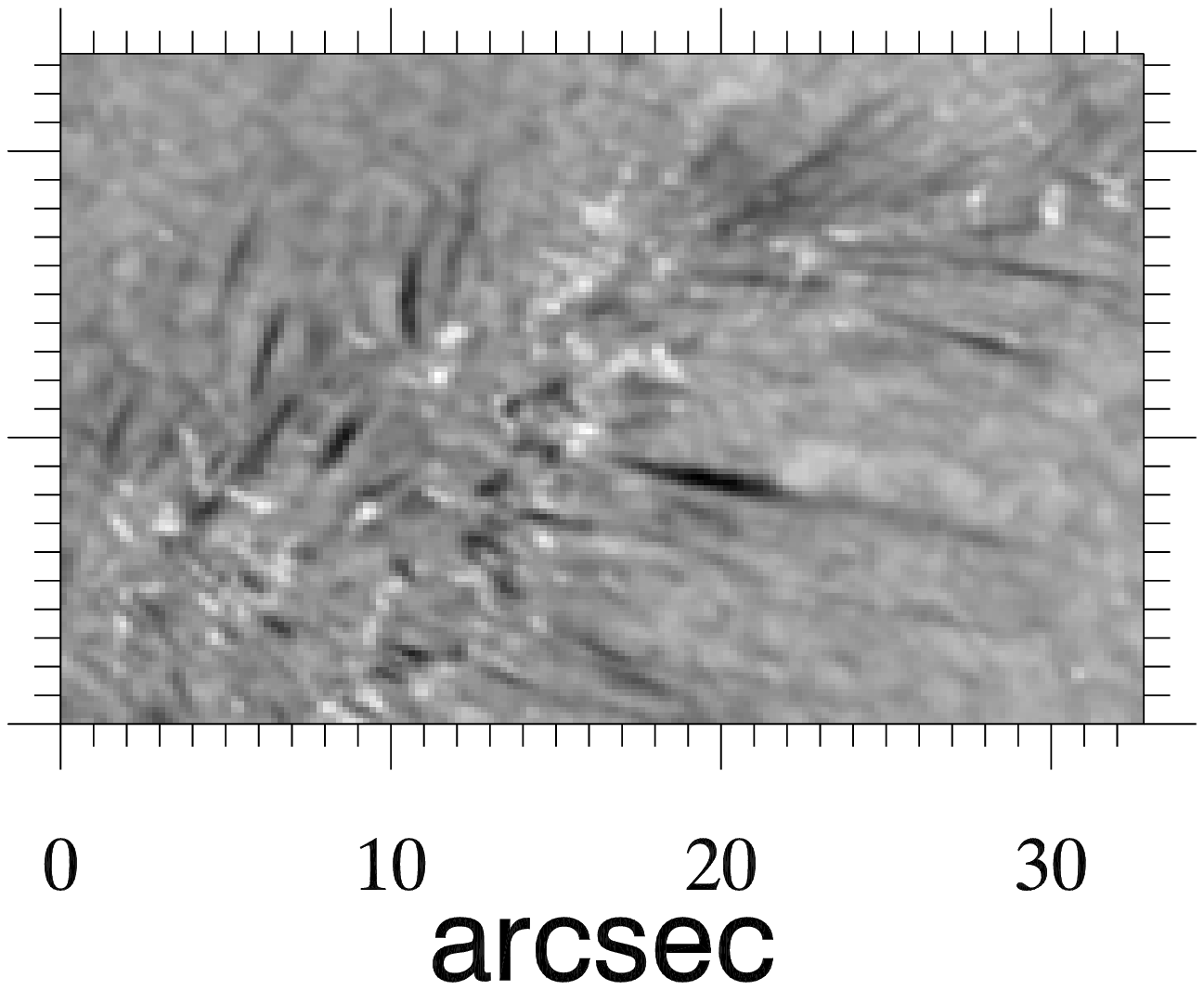}
\includegraphics[bb=100 61 475 400, clip,width=0.237\linewidth]{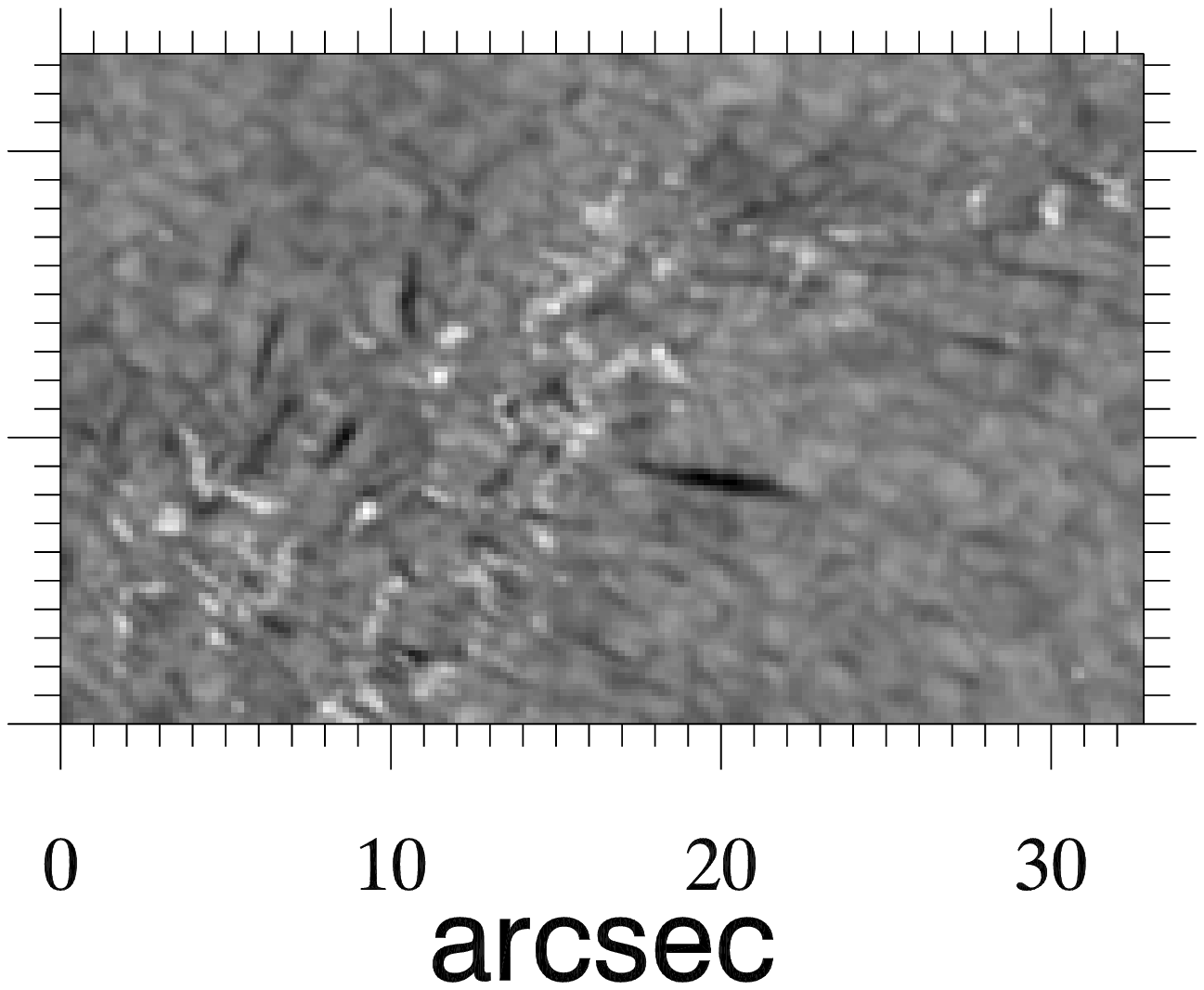}
\includegraphics[bb=100 61 475 400, clip,width=0.237\linewidth]{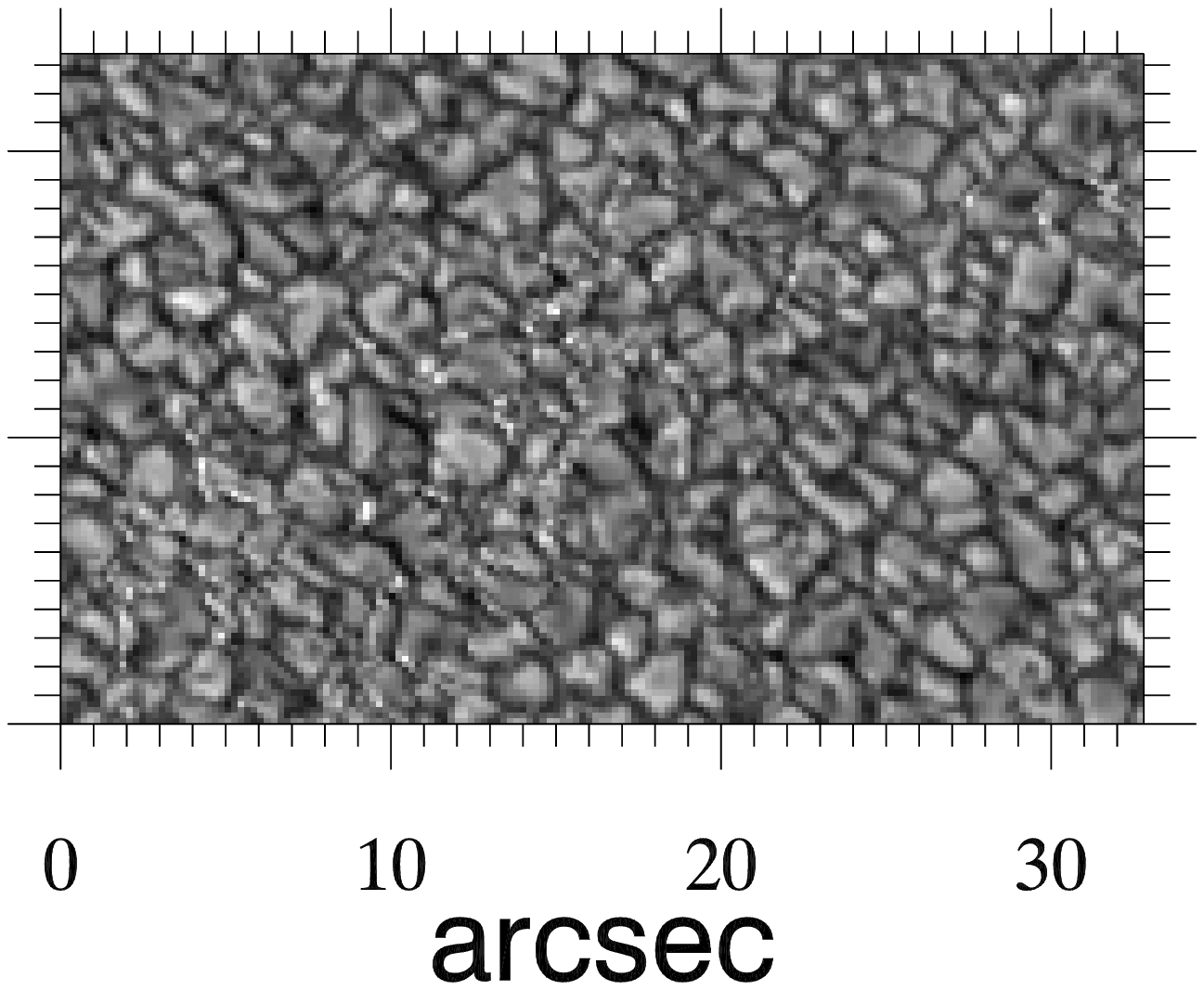}
\caption{A sample of speckle reconstructed narrow-band scan and the
  broad-band image ({\it lower right}) belonging to the same field of
  view. The size of each image is $32''\times\,23''.6$. The wavelength
  positions decrease in each image by 125\,m\AA~ starting from 6563.8\,
 \AA~ at the upper-left corner. The last image in the second row is
  the closest to the center of the line profile. }
\label{fig:scan}
\end{figure*}

Broad-band images were restored by using the spectral ratio (von der
L\"uhe 1984) and the speckle masking technique (Weigelt 1977).
Narrow-band images were reconstructed by using Keller and von der
L\"uhe's (1992) method.

As a result of this method, we obtained one reconstructed narrow-band
image for each wavelength position of the H$\alpha$ line.
Figure~\ref{fig:scan} displays the reconstructed narrow-band images
through one scan.  The observed region contains almost parallel
elongated dark mottles, which form the so-called 'chains of mottles'.

For  each pixel  in the  field of  view, H$\alpha$  line  profiles were
constructed from intensity values of narrow band images obtained at 18
wavelength positions.

\section{The Cloud Model}

The shapes and  amplitudes of stellar  spectral line profiles  reflect the
physical  properties  of  the  regions they  are  formed in. Therefore,
physical parameters such  as chemical abundance, density, temperature,
velocity, magnetic field, micro-turbulence, etc.  of these regions can
be inferred through the analysis of observed spectral line profiles.

The standard cloud model (Beckers, 1964) has been used extensively to
determine the physical properties of solar chromospheric fine
structures. The model considers a chromospheric feature like a cloud
above a uniform atmosphere described by a reference background
profile. The approach works well for optically thin structures and
gives estimates of source function $S$, optical thickness at the line
center $\tau_0$, Doppler width $\Delta\lambda_{\rm D}$, and
line-of-sight velocity $v_{\rm LOS}$ for the observed cloud.  These
parameters are assumed to be constant within the cloud along the line
of sight. The observed contrast profiles are matched with the
theoretical contrast profiles using the following formula

\begin{equation}
  C(\lambda)={{I(\lambda)-I_0(\lambda)}\over{I_0(\lambda)}}=\left({S\over{I_0(\lambda)}}-1\right)(1-e^{{-}\tau(\lambda)})
\end{equation}
where  I$_0(\lambda)$   is  the  reference  profile   emitted  by  the
background  and   $\tau(\lambda)$  is  the   optical  thickness.   The
wavelength dependence of optical thickness is given by

\begin{equation}
  \tau(\lambda)=\tau_0\exp\left[-\left({\lambda-\lambda_c(1-\upsilon_{LOS}/c)\over \Delta\lambda_D}\right)^2\right]
\end{equation}
where $\lambda_c$ is  the line center wavelength and  $c$ is the speed
of light. The standard cloud model considers chromospheric fine structures
to  be  completely  isolated   from  the  surrounding  atmosphere  and
illuminated from the underlying atmosphere.

The result of the cloud modeling is based on fitting the observed
contrast profile with the model from an iterative least square
procedure for non-linear functions.  The line core intensity and the
line core position of the observed profile were taken respectively as
initial values of $S$ and $\lambda_c$ for the iteration procedure.
For  $\tau_0$ and  $\Delta\lambda_{\rm D}$ the  values 1
and 0.3\,\AA~were used.

In order to determine the physical parameters of dark mottles 
using the cloud model, we selected dark features in the field of view
by masking pixels greater than 0.9 of mean value of the line
center intensity image for each spectral scan. Then we constructed
line profiles for each selected pixel and calibrated them relative 
to the quiet sun continuum. Finally, we calculated a contrast profile 
for each selected pixel using the mean of all profiles in the field of 
view as the background profile and applied the cloud model.

\section{Results and Discussion}
\subsection{Results of Cloud Modelling}
\begin{figure}
  \centering
 \includegraphics[width=0.45\linewidth]{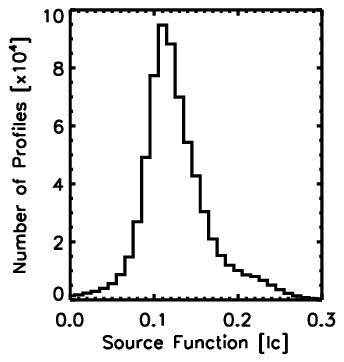}
 \includegraphics[width=0.45\linewidth]{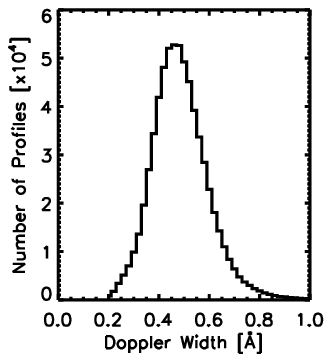}
\includegraphics[width=0.45\linewidth]{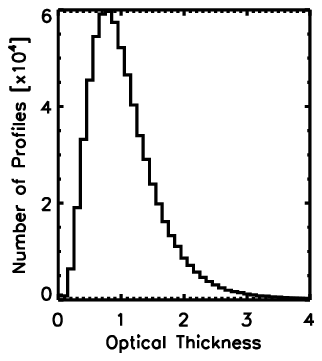}
 \includegraphics[width=0.45\linewidth]{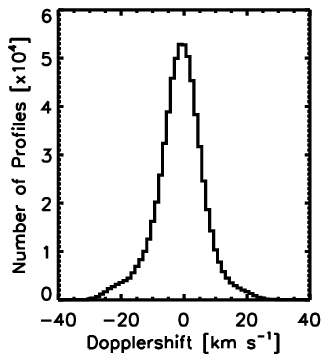}
 \caption{Distribution of cloud parameters. Negative values indicate
   a downward motion in the line-of-sight velocity.}
\label{fig:hist}
\end{figure}

Figure~\ref{fig:hist} displays histograms of cloud parameters 
  inferred for all the time series for dark mottles.  The source
function distribution shows a peak near 0.11 (in units of the
continuum intensity $I_c$), while the Doppler width's peak is close to
0.46\,\AA.  The peak of the optical thickness distribution is around
0.8, which indicates that dark mottles are mostly optically thin
structures.  The distribution of line-of-sight velocities varies
between $-$30 and 30\,km\,s$^{-1}$, and is almost symmetric, which
implies the presence of both downward and upward motions. It peaks
around $-$1.25\,km\,s$^{-1}$ (downward).

\begin{figure}
 \centering
 \includegraphics[width=0.49\linewidth]{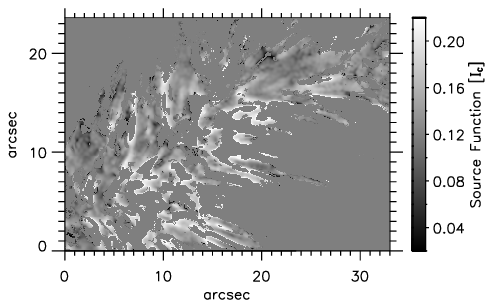}
  \includegraphics[width=0.49\linewidth]{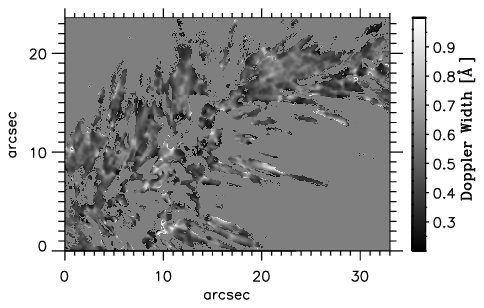}\\
 \includegraphics[width=0.49\linewidth]{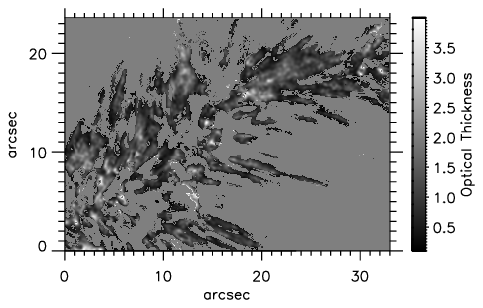}
 \includegraphics[width=0.49\linewidth]{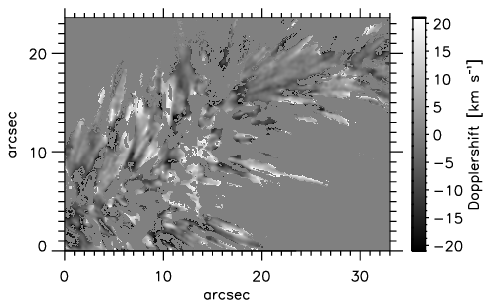}
 \caption{Maps of the source function, Doppler width, optical
   thickness at the line center and line of sight velocity. In the
   medium grey areas either the cloud model fit was not applied or did
   not give meaningful results at all. In the Doppler shift map, the
   bright and dark patches indicate upflows and downflows, respectively.}
 \label{fig:parimg}
 \end{figure}

Figure~\ref{fig:parimg}  shows  maps  of  the cloud  model  parameters
obtained for one  scan. The map of source  function has minimum values
at the center regions  of individual  dark mottles.  Through the  edge of
the structures, the source function values are close to the line center intensity
of the  background profile.  We observe that  regions of  high Doppler
width  values  have  low   optical  thickness  and  vice  versa.  This
anticorrelation was also noticed  by Alissandrakis  et al.  (1990) and
Tsiropoula et  al. (1993). On  the map of the Doppler shift, downflows
and upflows are always present  along mottles while material is mostly
descending near their footpoints.

 Table~\ref{tbl:darkcloudcomp} shows values of the cloud parameters
 reported by various authors.  It can be seen from the table that our
 results, inferred for all the time series are in good agreement
 with previous measurements.
  Several intrinsic parameters can be determined by using the
   cloud model parameters. Assuming 15\,km\,s$^{-1}$ for
   microturbulent velocity, $\xi_t$, we can calculate temperature from
   the deduced Doppler width values;
\begin{equation}
\Delta\lambda_{D}={\lambda_0\over c}\sqrt{{2kT\over m_H}{+\xi^2_t}}
\end{equation}

The relation between the optical thickness at the line center,
the Doppler width and the number density in the
second hydrogen level $N_{2}$ may be given:

\begin{equation}
\tau_0={\pi^{1\over{2}}e^2\over{m_ec}}{f\lambda^2\over{c}}{N_{2}L\over\Delta\lambda_D}{d}
\end{equation}
 inserting the constants in the equation above, we obtain: 

\begin{equation}
\tau_0=1.38\times10^{21}{N_{2}d\over\Delta\lambda_D}
\end{equation}

The thickness of mottles estimated at the images was assumed to
$\approx$350 km, considering a cylindrical structure, and the
inclination was taken as $\approx$30$^{\circ}$ given for spicules by Heristchi
and Mouradian (1992).  Therefore the geometrical width, $d$ of the
structures under investigation, is equal to $\approx$700 km. Then using
the same method as Tsiropoula \& Schmieder (1997), we also obtained
the electron density $N_{\rm e}$, the total particle density $N_{\rm
  H}$, the gas pressure $p$, the total column mass $M$, the mass
density $\rho$, and the degree of hydrogen ionization
$\chi_{\rm{H}}$. The relations can be expressed as:

\begin{equation}
N_{\rm e}=3.2~10^8\sqrt{N_2}
\end{equation}
\begin{equation}
N_{\rm H}=5~10^8~10^{0.5~\log N_2}
\end{equation}
\begin{equation}
p=k(N_{\rm e}+1.0851~N_{\rm H})~T
\end{equation}
\begin{equation}
M=(N_{\rm H} m_{\rm H}+0.0851~N_{\rm H}\times3.97m_{\rm H})~d
\end{equation}
\begin{equation}
\rho={M\over d}
\end{equation}
and
\begin{equation}
\chi_{\rm H}={N_{\rm e}\over N_{\rm H}}
\end{equation}

 We assumed histograms of the parameters to be normally
  distributed, so we fit them with a Gaussian curve separately. We
  took peak values of Gaussian distributions as a mean value and their
  1$\sigma$ widths as a standard error.  In Table~\ref{tbl:darklit},
we present the results obtained from all the time series and
compare these with the values found by Tsiropoula \& Schmieder (1997)
and Tsiropoula \& Tziotziou (2004) for dark mottles.  The results are
  generally in agreement with previous computations, although some
  variations do exist.  These differences may arise from various
  factors. First, the atmospheric seeing when the observation was
  performed, has a non-negligible effect on the determination of cloud
  parameters, especially on the Dopplerwidth (Tziotziou et
  al.,~2007). Therefore, it causes an uncertainty in the calculation
  of the pressure and temperature, which depends on the Dopplerwidth
  and the microturbulent velocity, which is assumed.  Second, any
  uncertainties raised at the determination of the geometrical
  thickness of the structures, $d$, will be propagated to the
  uncertainties in the calculations of $N_{2}$, $N_{\rm e}$ and $M$.
  Finally, differences in the inferred values of $\tau_0$ can cause
  variations in these parameters. 


\begin{table*}
\centering
\caption[]{
Cloud model parameters for dark mottles given by different authors.}
\begin{tabular}{lcccc}
\hline
 Authors& $S$&$\upsilon$ (km~s$^{-1}$)&$\tau_0$&$\Delta\lambda_D$
 (\AA )\\
\hline
Beckers 1968& 123 &-&1.4&0.5 \\
Bray 1973&130 to 160&-9 to 7&1.0&0.5\\
Grossmann-Doerth $\&$&130&-8 to 8&1.1&0.45\\
von Uexk\"ull 1977\\
Tsiropoula et~al. 1993&163.3$\pm14.3$& -0.26$\pm6.6$ &1.8$\pm1.1$&0.37$\pm0.1$\\
Lee et~al. 2000&0.16~$I_c$&-2.8&2.2$\pm0.5$&0.55\\
Tziotziou et~al. 2003&0.15~ $I_c$&-0.1&0.9&0.35\\
Al et~al. 2004&0.14~ $I_c$&0.18&1.58&0.44\\
This work&0.11~$I_c$&-1.25&0.8&0.46\\
\hline
\end{tabular}
\label{tbl:darkcloudcomp} 
\end{table*}

\begin{table*}
\centering
\caption[]{
Pyhsical parameters for dark mottles given by different authors.}
\begin{tabular}{lccc}
\hline
Parameters& Tsiropoula \& Schmieder& Tsiropoula \& Tziotziou& This work\\
&(1997)&(2004)&\\
\hline
$N_1 (cm^{-3})$&$(1.6\pm0.8) 10^{10}$&$-$&$(3.0\pm0.7) 10^{10}$\\
$N_2 (cm^{-3})$&$(1.4\pm1.1) 10^{4}$&$(4.2\pm2.0) 10^{4}$&$(4.2\pm1.9) 10^{4}$\\
$N_e (cm^{-3})$&$(3.4\pm1.5) 10^{10}$&$(6.4\pm1.6) 10^{10}$&$(7.0\pm1.6) 10^{10}$\\
$N_H (cm^{-3})$&$(5.1\pm2.1) 10^{10}$&$(9.9\pm2.5) 10^{10}$&$(10.8\pm2.5) 10^{10}$\\
$T (K)$&$(1.0\pm0.8 ) 10^{4}$&$(1.0\pm0.3) 10^{4}$&$(1.2\pm0.9) 10^{4}$\\
$P (dyn~cm^{-2})$&$0.15\pm0.1$&$0.24\pm0.1$&$0.32\pm0.2$\\
$M ~(gr~cm^{-2}$)&$(2.2\pm0.4) 10^{-5}$&$(2.2\pm0.6) 10^{-5}$&$(1.6\pm0.4) 10^{-5}$\\
$\rho~(gr~cm^{-3}$)&$(1.1\pm0.5) 10^{-13}$&$(2.2\pm0.6) 10^{-13}$&$(2.4\pm0.6) 10^{-13}$\\
$\chi_{\rm{H}}$&0.65$\pm$0.1&0.65&$0.64\pm0.1$\\
\hline
\end{tabular}
\label{tbl:darklit}
\end{table*}


We calculated the pressure inside the dark mottles as
0.32\,dyn\,cm$^{-2}$. This result is in agreement with
Heinzel \& Schimieder (1994), who concluded that classical cloud model can
be applied to low pressure structures ($<$\,0.5\,dyn\,cm$^{-2}$),
assuming a constant source function and a rather low opacity.

Previous studies of spicules show that they have temperatures of about
5\,000-15\,000\,K, and densities of about
3$\times$10$^{-13}$\,g\,cm$^{-3}$.  The physical properties we
inferred from cloud modeling here for the mottles are in good
agreement with the values found earlier for the spicules showing
further evidence that these two features have significant similarities
(Table~\ref{tbl:darklit}).

\subsection{Temporal Analysis of a Mottle}

\begin{figure}
 \centering
 \includegraphics[width=0.95\linewidth]{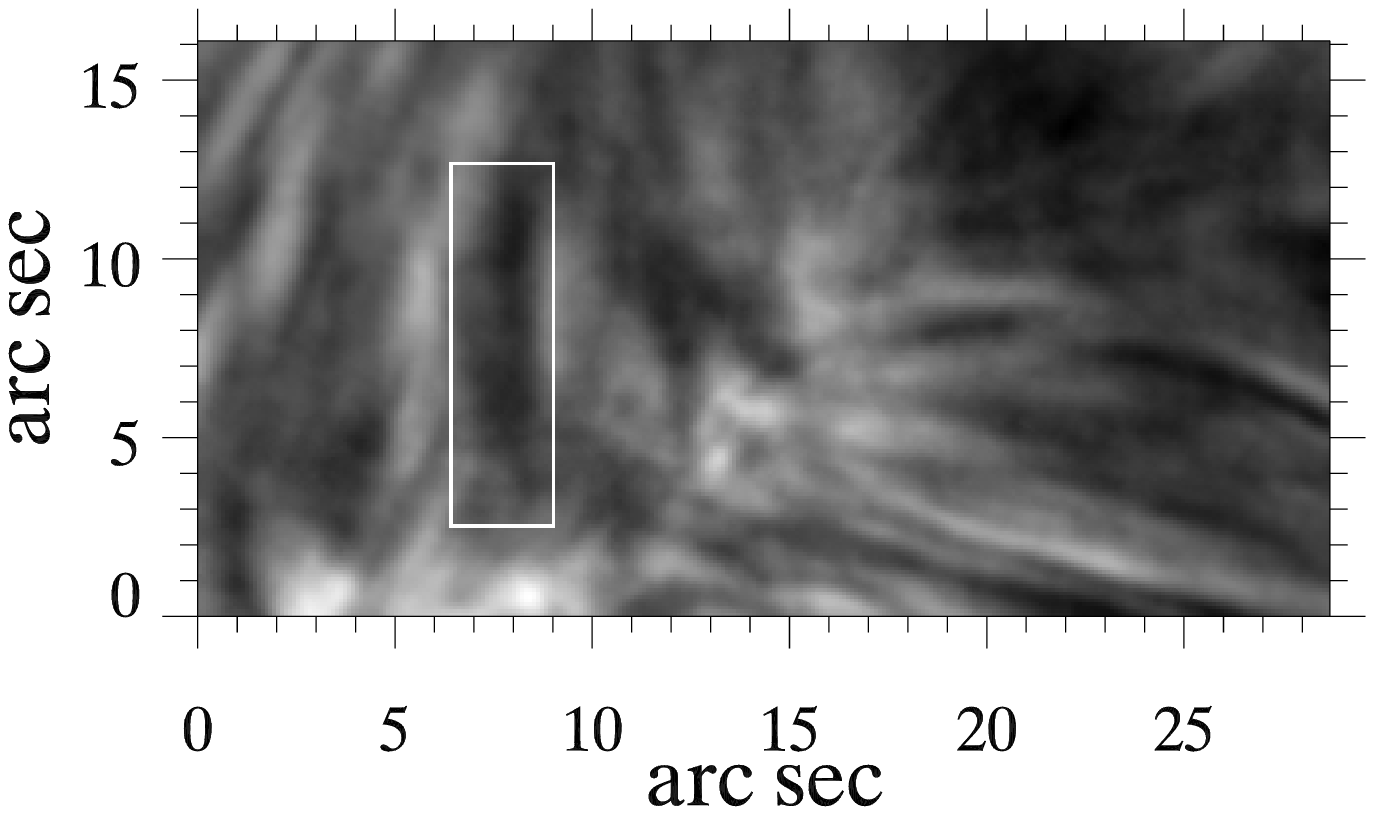}
 \includegraphics[width=0.95\linewidth]{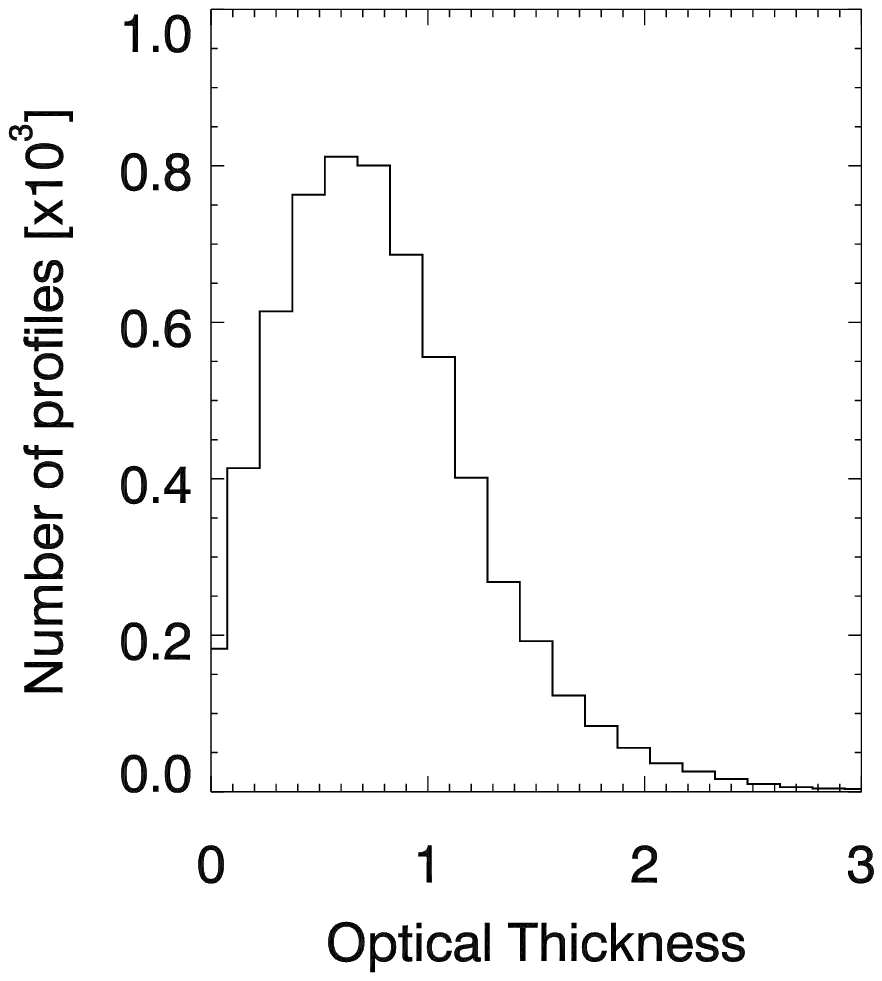}
 \caption{The mean image of H$\alpha$ line center intensity series
   ({\it top}). White rectangle inside the image marks the mottle
   selected for the Fourier analysis.  The distribution of optical
   thickness for the mottle subjected to timing analysis. ({\it
     bottom}). }
 \label{fig:histtau0}
 \end{figure}

In order to study the oscillatory behaviour of chromospheric features,
we performed a Fourier analysis to the fluctuations of cloud model's
source function and Dopplershift for a long horizontal mottle's
region.  This region  which is centered at (8\arcsec, 8\arcsec)
is shown with a white rectangle in Figure~\ref{fig:histtau0}. We
selected this region because it could be well isolated from other
structures and the cloud model fit it well for all of the time series.

\begin{figure}
 \centering
 \includegraphics[width=0.95\linewidth]{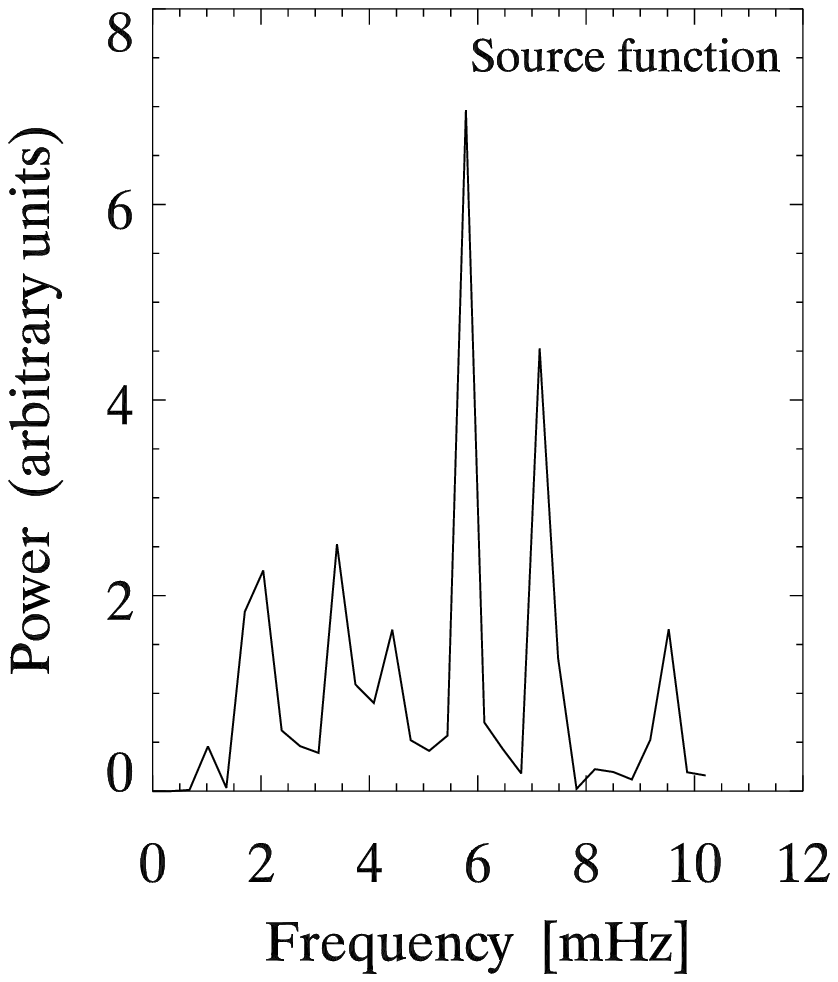}
 \includegraphics[width=0.95\linewidth]{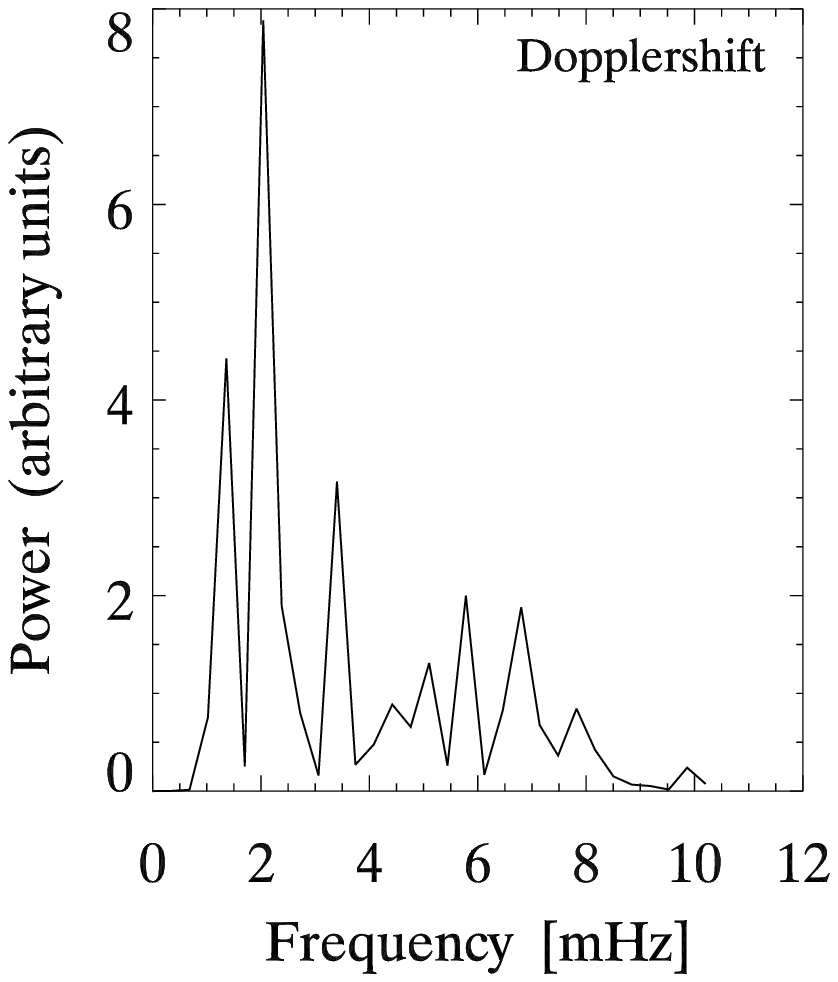}
 \caption{ Power spectra for the cloud
   source function and Dopplershift fluctuations of the mottle ({\it
     top and bottom, respectively}). }
 \label{fig:power}
 \end{figure}

 Before computing the power spectra, we averaged the signal in
   the selected mottle's region and then applied a polynomial trend
 removal of the 4th order to the time series of source function and
 velocity. Consequently, we cut down the power at frequencies lower
 than 0.7\,mHz, which corresponds to half of the total duration of our
 observation. This procedure also affected the amplitude of the power
 at the frequencies lower than 1.3\,mHz, but the power at high
 frequencies remained unaffected.

 In Figure~\ref{fig:power}, we show the power spectra of Cloud
 Dopplershift and source function variations in the considered
 mottle's region. Power distribution of source function peaks at high
 frequencies between 5 and 8\,mHz with a maximum around 5.7\,mHz (i.e
 periods of 3 minutes) which corresponds to the acoustic cut-off
   frequency. The second significant power with a lower amplitude is
 around 7.2\,mHz. De Pontieu et al. (2007) also reported a significant
 3 minute periodicity in the intensity variations of the overlying
 loops emanating from a network. They considered that the lower
 opacity of these structures allows glimpses of internetwork dynamics
 and oscillations underneath. In order to check this scenario, we
 looked into the distribution of optical thickness for the considered
 mottle (Figure~\ref{fig:histtau0}). The peak is around 0.6, which is
 smaller than the peak of general distribution of optical thickness
 presented in Figure~\ref{fig:hist}. This shows that the long mottle
 under consideration has a lower opacity and also confirms the
 suggestion of De Pontieu et al. (2007). In the power spectrum of
 source function, significantly lower powers are seen at low
 frequencies around 2.2 and 3.5\,mHz.

 The cloud Dopplershift power spectrum shows three well separated
 peaks at lower frequencies. The most significant peak lies around
 2.2\,mHz, i.e. at periods of 7 minutes. The other two are around
 1.3\,mHz and 3.5\,mHz (i.e periods of 5 minutes). We should note that
 magnitude of the power around 1.3\,mHz is strongly affected by the
 procedure of trend removal, so we can not form a conclusion about its
 relevance. Also, peaks in the 3 minute range are visible but they
 have remarkably less significance.

Similar to our results for the periodicities in velocity variations,
Tziotziou et al. (2004) found a dominant period in the 5 minutes
(3.3\,mHz) for intensity and velocity variation of dark mottles. De
Pontieu et al. (2007) reported that the mottles are dominated by
oscillatory behavior with periods around 5 minutes and longer. More
recently, Tsiropoula et al. (2009) found a significant peak around the 5
minute range for the mottles' region.

 This strong peak around 5 minutes which was observed in the
  Doppler velocity power spectrum has long been attributed to p-mode
  oscillations generated at the photosphere propagate upward in and
  around magnetic flux concentrations of a network (Giovanelli et al.
  1978).  Several studies based on simulations and observations have
  suggested that low frequency ($<$5~mHz) magnetoacoustic waves might
  propagate into the chromosphere along magnetic field lines which are
  significantly inclined with respect to the surface gravity (Bel \&
  Leroy 1977; Suematsu 1990; De Pontieu et al.  2004; Jefferies et
  al. 2006).   In order to study the propagation characteristics of
waves at different heights in atmosphere, it is essential to compute
phase difference and coherence spectra.  In a future paper, we will
concentrate on this issue in detail using the lambdameter method
(Tsiropoula et al., 1993; Al et al., 2004) to compute intensity and
Dopplershift image series for different widths through the H$\alpha$
profile.

In   this   work,  using   high   spectral   and  spatial   resolution
spectro-imaging observations,  we constructed H$\alpha$  line profiles
for every pixel of mottles in  the field of view. We fit the resulting
profiles with the  cloud model and infered physical  parameters of the
dark mottles, which agree with  previous studies.  We also performed a
Fourier  analysis on  fluctuations of  the cloud  velocity  and source
function to detect the  oscillatory behaviour of the mottles. However,
we could only concentrate on  one mottle  in the field  of view  since the
cloud model can  not fit the all of the  profiles.  Our analysis shows
that the mottle is dominated by 3 minute periods (the acoustic cut-off
period  in the  chromosphere) in  source  function, and  periods of  5
minute and longer in Dopplershift.

\acknowledgements I thank the anonymous referee for valuable comments
and suggestions. I am indebted to Nurol Al Erdogan for supplying the
observation and for careful reading of the manuscript. This work was
supported by Scientific Research Projects Coordination Unit of
Istanbul University. Project number 851.  This work was partially
supported by Deut\-sche For\-schungs\-ge\-mein\-schaft through grant
Kn 152/26-1. The Vacuum Tower Telescope is operated by the
Kiepenheuer-Institut f\"ur Sonnenphysik, Freiburg, at the Spanish
Observatorio del Teide of the Instituto de Astrof$\acute{\i}$sica de
Canarias.

\newpage

\end{document}